# Radio Transmission Performance of EPCglobal Gen-2 RFID System

Manar Mohaisen, HeeSeok Yoon, and KyungHi Chang
The Graduate School of Information Technology & Telecommunications
INHA University
Incheon, Korea
lemanar@hotmail.com, electech@empal.com, khchang@inha.ac.kr

*Abstract* — In this paper, we analyze the performance of the encoding and the modulation processes in the downlink and uplink of the EPCglobal Gen2 system through the analysis and simulation. Furthermore, the synchronization issues on time and frequency domain and the preamble architecture are evaluated. Through the simulation in the uplink, we find that the detection probability of FM0 and Miller coding approaches 1 at 13dB $E_b/N_0$.

## 1. Introduction

RFID (Radio Frequency Identification) is a technology where the reader accesses automatically the data stored on the tag, label, or card, which have built in a micro-chip.

EPCglobal Class-1 Generation-2 standard is a passive RFID system where the reader transmits continuous wave (CW) to power on tags. This means that a tag does not have any energy source. Reader starts to communicate while tag responds to the reader with specified answers including random numbers and some identifying numbers.

EPCglobal Class-0 and Class-1 Standards have different protocols. For instance, in the downlink, Class-0 has three data symbols ('0', '1' and 'Null') with varying duty cycles. But Class-1 uses just two data symbols, Data '0' and '1'. Class-1 specifies two modes of operation, a fast mode for North America and a slower one for Europe. Class-1 Gen2 takes advantages of the two above classes. The off pulse of Gen2 has a fixed duration. Like Class-1, Gen2 uses just two data symbols, Data '0' and Data '1'.

It is difficult for Class-0 readers and Class-1 readers to address a specific tag. Additionally, there is a problem of collision between readers due to the large offset between the reader signal and the tag signal. In Class-1 Gen2, a reader can address single tag, so these problems are solved [1].

In this paper, we explain EPCglobal Class-1 Generation-2 RFID standard and then analyze the performance of wireless communications between tag and reader in EPCglobal RFID system based on EPC Gen 2 Specification. The remainder of the paper is as follows: In section 2, we assess the waveform through the simulation by applying encoding method and modulation method used in reader to tag communications. In section 3, we introduce the encoding and backscatter modulation process and simulate backscatter modulation used in tag to reader communications. In section 4, the importance and the role of frequency and time synchronization such as preamble and frame synchronization in the downlink and uplink are explained. In the last section 5, we draw some conclusions.

## 2. Reader to Tag Communications

Figure 1 shows the block diagram of the reader (transmitter) to the tag (receiver) communications. Only AWGN channel is considered in this section.

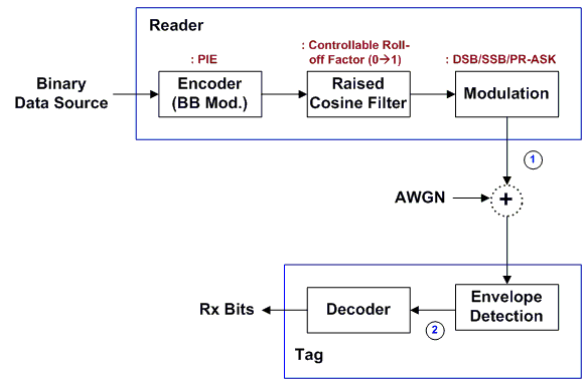

Figure 1. Reader to tag communications

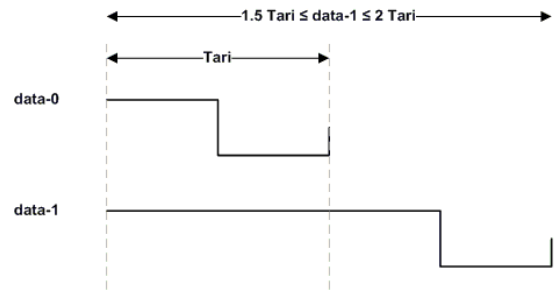

Figure 2. PIE symbols

## 2.1 Data Encoding

Reader uses Pulse Interval Encoding (PIE). The length of Data-0 is 1 Tari, where Tari is the time reference unit of the signaling and takes on values given by (1),

$$6.25\,\mu s \leq Tari \leq 25\,\mu s \quad (1)$$

Data-1 length can vary between 1.5 and 2 *Tari,* and Fig. 2 shows the symbols used in PIE encoding.

## 2.2 Modulation

Reader can use double side band - amplitude shift keying (DSB-ASK), Phase Reversal ASK (PR-ASK) and Single Side Band ASK (SSB-ASK) modulation schemes. In this section, we introduce these modulation schemes and the way to obtain the standard RF pulse shapes.

Figure 3 shows the standard forms of ASK and PR-ASK modulation. In the view of modulation depth in reader to tag communications, minimum value and typical value of modulation depth are 80% and 90%, respectively, where the modulation depth is given by (A-B)/A [1].

### 2.2.1 DSB-ASK Modulation

DSB-ASK is the conventional ASK modulation where the carrier is modulated by a modulating waveform obtained by encoding the data. Table 1 shows the important parameters used to obtain modulated carrier with raised cosine (RC) filter in Fig. 4 (a) and to detect the envelope by applying Hilbert transform in Fig. 4 (b).

### 2.2.2 PR-ASK Modulation

PR-ASK is obtained by inverting the phase of the baseband waveform at the boundary between symbols. RC filter with roll-off factor of 0.99 is also used and the envelope is detected using Hilbert transform. Fig. 4 (c) and (d) show the PR-ASK modulated carrier and detected envelope using Hilbert transform, respectively.

### 2.2.3 SSB-ASK Modulation

SSB-ASK Modulation uses the Hilbert transform to suppress one of the signal PSD two sides [2].

## 3. Tag to Reader Communications

Figure 5 shows the block diagram of the communication link from the tag (transmitter) to the reader (receiver) under AWGN channel.

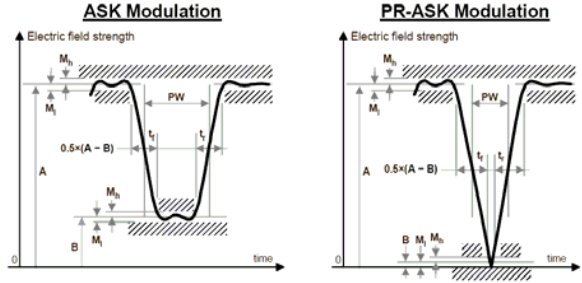

**Figure 3. Reader to tag RF envelope**

**Table 1. Simulation Parameters (DSB-ASK)**

| Parameter | Value |
| --- | --- |
| Data | [0 1 0 1 1 0] |
| Filter | Raised Cosine |
| Alpha (Roll-off Factor) | 0.99 |
| Envelope Detection | Hilbert Transform |

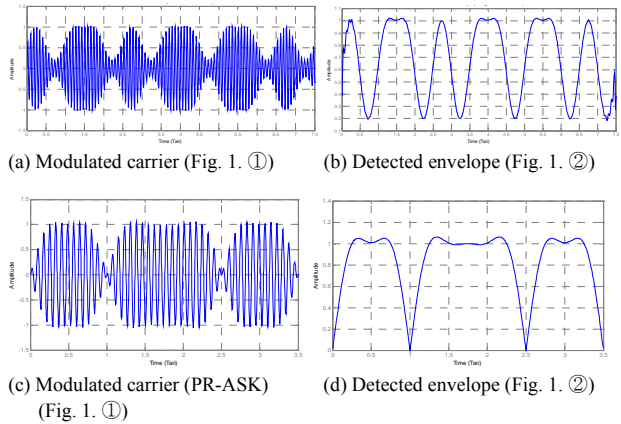

(a) Modulated carrier (Fig. 1. ①)   (b) Detected envelope (Fig. 1. ②)

(c) Modulated carrier (PR-ASK) (Fig. 1. ①)   (d) Detected envelope (Fig. 1. ②)

**Figure 4. Simulation results**

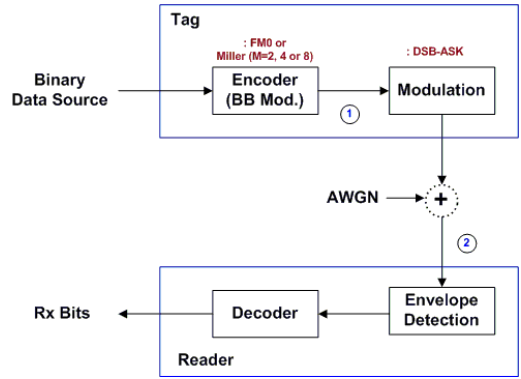

**Figure 5. Tag to reader communications**

### 3.1 Data Encoding

Tag communicates with reader using either FM0 or Miller sub-carrier encoding. The basis functions of these two encoding methods are the same, so, the BER performance is equal. On the other, Miller code can be spread to reduce the rate by multiplying the encoded symbols by a sequence. These sequences can include 2, 4, or 8 cycles per encoded symbol.

### 3.1.1 FM0 Encoding

A phase inversion occurs at the boundary between symbols while Data-0 has a mid-symbol phase inversion.

### 3.1.2 Miller Code

Miller code can use some sequences called 'Miller sub-carrier sequences', which reduces the transmission rate. 2, 4, or 8 cycles can be contained in one bit.

The transmission rate (or link frequency (*LF*)) varies depending on the parameters sent by the reader to the tag and on the encoding type used. The range of *LF* is given as following;

$$5 kbps \leq LF \leq 640 kbps \qquad (2)$$

Data are transmitted at the lowest data rate of 5 kbps when we use Miller code with $M = 8$, *TRcal* = 200 μs and *DR* = 8 (where M=2k is the symbol size, k is the number of bits and *DR* is the divide ratio and takes on one of two values; namely 64/3 and 8). The maximum data rate of 640 kbps can be achieved using FM0 encoding, *TRcal* = 33.3 μs and *DR* = 64/3. Data rate and *TRcal* are given by (3) and (4), respectively.

$$Data\ Rate = LF = \frac{DR}{TRcal \times M} \qquad (3)$$

$$1.1 \times RTcal \leq TRcal \leq 3 \times RTcal \qquad (4)$$

### 3.2 Backscatter Modulation

The working principle of backscatter modulation is as follows; Reader transmits a continuous wave (CW), this CW is rectified at the tag and used as a power supply. As a consequence, the tag turns on [3]-[5]. In EPC Gen-2, the ASK type (modulates the amplitude) and the PSK type (modulates the phase) can be used [1]. Table 2 lists the simulation parameters.

The received signal backscattered from the tag is mixed with CW used in the transmitter. The output of the mixer is low-pass filtered using lowpass filter to remove the high frequency component. The resulting signal is then decoded according to the code used at the tag transmitter.

Figure 6 (a) depicts the baseband signal of the encoded data sequence [1 1 0 0 1 0 0 1]. Figure 6 (b) shows the backscattered signal from the tag. Figure 7 shows the signal at the output of the mixer.

**Table 2. Simulation Parameters (Backscatter Mod.)**

| Parameter | Value |
|---|---|
| Reader Tx | CW (Sine Wave) |
| CW Frequency | 896 MHz |
| Sampling Frequency | 5 × 896 MHz |
| Tx Bits | 8 Random Bits |
| Rx Lowpass Filter (LPF) | Butterworth 1st Order |
| Tag Coding | FM0 |
| Tag Modulation | ASK (100 %) |
| Tag Data Rate | 320 Kbps |

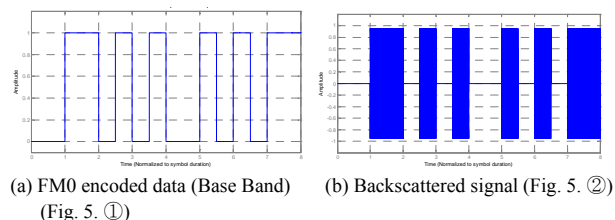

(a) FM0 encoded data (Base Band) (Fig. 5. ①)  (b) Backscattered signal (Fig. 5. ②)

**Figure 6. Simulation results**

### 3.3 BER Performance of FM0/Miller Code

In the encoding methods used in UL (from tag to reader), namely FM0 and Miller encoding, the same basis functions are used. As a consequence, the BER performance will be the same, if we apply a symbol-by-symbol detection. Note that the symbol-by-symbol detection method is not optimal, but easy to implement, compared with differential detection.

The decision rule at the receiver is given by (5) [6]-[7],

$$\begin{cases} Choose\ H_1,\ if\ \left( \int_0^{T/2} \tilde{r}(t)dt \cdot \int_{T/2}^{T} \tilde{r}(t)dt \right) \geq 0. \\ Choose\ H_0,\ otherwise. \end{cases} \qquad (5)$$

where Data-1 is represented by $H_1$, and Data-0 by $H_0$. $\tilde{r}$ is the base-band received signal, and *T* is the symbol duration. Then, the symbol error rate (SER) (or equivalently the BER) is given by (6) [8],

$$P_e = 2Q\left(\sqrt{\frac{E}{N_0}}\right)\left[1 - Q\left(\sqrt{\frac{E}{N_0}}\right)\right] \qquad (6)$$

where *E* is the symbol energy and $N_0/2$ is the PSD of the complex double-sided AWGN.

Figure 8 shows the BER performance of the FM0 code. $E_b/N_0$ of about 13dB is required to meet the target BER of $10^{-3}$. The simulation result is identical to the theoretical result.

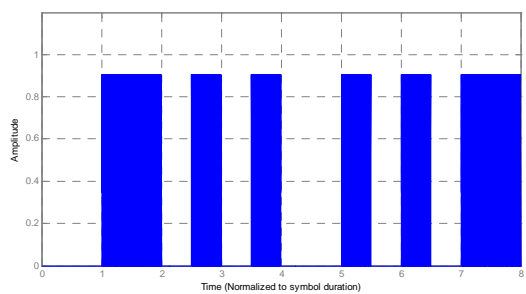

**Figure 7. Signal at the output of the mixer**

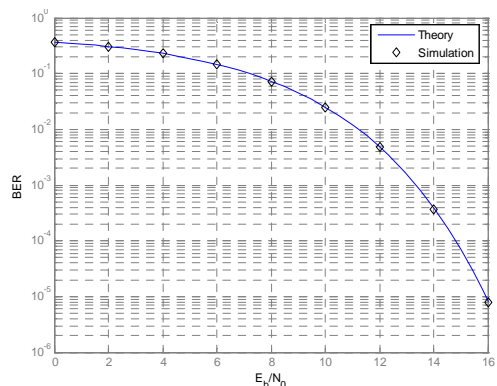

**Figure 8. FM0 BER performance**

## 4. Frequency and Time Synchronization

Before decoding the received data, transmitter frequency and timing must be synchronized. Nevertheless, the way of synchronizing the transmitter and receiver depends on the structure and the technology used in the communication systems. In the following sections, we introduce the principle and way of obtaining synchronization in RFID system.

### 4.1 Downlink Synchronization
#### 4.1.1 Frequency Synchronization
In RFID system, tags use backscattering modulation. This means tags don't have local VCO (Voltage Controlled Oscillator) to generate a local carrier. The received CW is scattered back to the reader after being modulated by the encoded bits. The first step in the demodulation process in the DL is to suppress the carrier by detecting the envelope. This is why frequency synchronization is not needed in the DL (from reader to tag).

#### 4.1.2 Time Synchronization
EPCglobal standard uses PIE encoding to transmit data in the DL. The reason why this encoding method is used is that it implicitly includes the clock to detect the bit (or symbol) boundaries easily at the tag with a small hardware. The clock is used to make the interaction between the analog part (analog front end (AFE) and modulator etc.) of the tag and the digital part (i.e., EPROM and control part).

Figure 9 shows the functional block diagram of the tag's demodulator [9]. We use the Hilbert transform followed by a low-pass filter to obtain the signal envelope. An analytic signal is a complex signal, which is given by (7),

$$g_+(t) = g(t) + j\hat{g}(t) \quad (7)$$

where the real part is the original signal and the imaginary part is the Hilbert transform of the original signal. The original signal is time-delayed before being added to the Hilbert transform to match the delay caused by the Hilbert transform, which is one-half the length of the Hilbert filter. The envelope of the signal can thus be found by taking the absolute value of the analytic signal. In order to eliminate ringing and to smooth the envelope, we apply a low-pass filter.

At first, the carrier is suppressed by detecting the envelope (base-band signal). The envelope of the received signal is the encoded symbols. The envelope is passed through a trigger which detects the rising and falling edges of the envelope (its output is changed accordingly) and cleans it from the noise.

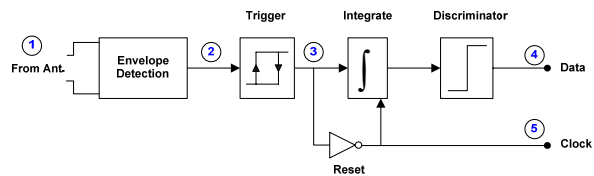

**Figure 9. Functional block diagram of tag's demodulator**

**Table 3. Simulation Parameters (DL Timing Synch.)**

| Parameter | Value |
| --- | --- |
| Carrier Frequency | 910 MHz |
| Reader Encoding | PIE |
| Reader Modulation | ASK |
| Transmitted Bits | 1 0 0 1 1 0 |
| '1' / '0' Duration | 1.5 / 1 Tari |
| Envelope Detection | Hilbert Transform |

The output of the trigger is inverted to get the clock which is used in the digital part of the tag and to reset the integrator which integrates the output of the trigger. The output of the integrator is passed through the discriminator which decides if the transmitted bit was '1' or '0'. The data is passed through the digital part in-parallel with the clock to be further processed.

In figure 10, we evaluate the functional performance of the tag receiver by tracing the signal at the output of every functional block in Figure 9. The simulation parameters for time synchronization in the DL are shown in Table 3.

Tag discriminates between bits using the pivot value, where the pivot is defined as

$$Pivot = \frac{RTcal}{2} \quad \text{with} \quad 2.5 \times Tari \leq RTcal \leq 3.0 \times Tari \quad (8)$$

Received bit with high level duration less than the pivot value is considered as bit '0'. Otherwise, it is considered to be bit '1'.

As one can see from the above Figure 10, the received signal envelope can be easily detected. The clock is sent to the data symbols to set the integrator and to trigger the memory to receive detected bits.

Falling and rising edges can be detected from the trigger output or the clock. Figure 11 depicts that the falling and rising edges can be detected by differentiating the clock signal depicted in Figure 10 (d). A positive value means a falling edge of the trigger output which represents the change from high level to low level in the same symbol. A negative value indicates a falling edge in the trigger output which represents the boundary between two symbols (bits).

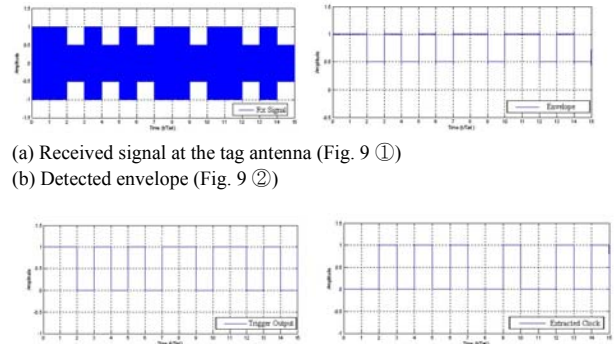

(a) Received signal at the tag antenna (Fig. 9 ①)
(b) Detected envelope (Fig. 9 ②)

(c) Output of the trigger (Fig. 9 ③)
(d) Extracted clock (Fig. 9 ⑤)

**Figure 10. Simulation results**

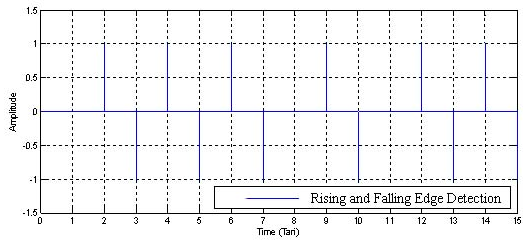

**Figure 11. Falling and rising edges detection from the clock signal**

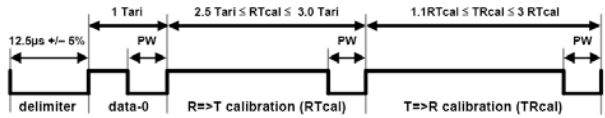

**Figure 12. Reader to tag preamble**

We use the rising edge indexes to reset the integrator, and the integrator outputs are fed to the discriminator. Based on the received signal from integrator, the discriminator decides if the transmitted bit was '1' or '0'.

*4.1.3 Reader Preamble and Frame Synchronization*

Every reader signaling form must begin with a preamble or a frame-synchronization. Query command must begin with a preamble which indicates the beginning of the inventory of tags while other commands begin with a frame-synchronization. Fig. 12 represents the preamble using PIE encoding.

The preamble starts with a delimiter followed by Data-0. The Data-0 is followed by the *RTcal* (Reader to Tag calibration) used to distinguish between Data-0 and Data-1. The pivot is the threshold of comparison. The symbol with duration shorter than the pivot is considered to be data-0. Otherwise, it is considered to be data-1. *TRcal* (Tag → Reader Calibration) is given by (4) to calculate the Link Frequency (*LF*).

*DR* is sent from the reader to the tag using Query command using one bit. If the bit '0' is transmitted, *DR* takes on the value '8'. Otherwise, it takes on the value '64/3'. The value of pivot can be changed only by transmitting a Query command. The minimum duration between consecutive Query commands is 8 *TRcal*.

The signaling for frame synchronization has the same structure as that of the preamble except for the absence of the *TRcal*.

### 4.2 Uplink Synchronization
*4.2.1 Frequency Synchronization*

Frequency synchronization is not considered in RFID system because at first stage, reader suppresses the carrier and detects the envelope.

*4.2.2 Time Synchronization*

The clock is extracted from the base band signal itself as indicated in the previous section. For every encoding method, there is a suitable clock extraction method.

*4.2.3 Detection Probability for Preamble*

We consider error to occur independently. As a consequence, the probability of receiving a fully error-free preamble is given by multiplying the individual symbols' probabilities of correct reception. The probability of correct detection of the preamble is thus given by (9),

$$P_{cp} = (1 - P_e)^N, \quad (9)$$

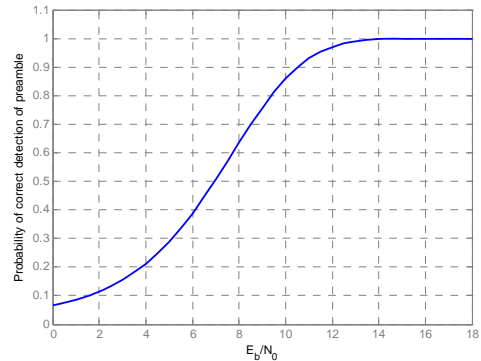

**Figure 13. UL Preamble's probability of correct detection**

where $P_e$ is the probability of symbol error and $N$ is the number of symbol included in the preamble. We consider herein that the preamble includes 6 symbols. As we commented before, the BER of FM0 and Miller (M=1) are the same. So, detection probability for FM0 Preamble is equivalent to that of Miller code.

From Figure 13, we confirm that probability of detection for preamble approaches 1 at $E_b/N_0$ of 13dB.

## 5. Conclusions

In this paper, we simulate the performance of wireless communications in EPCglobal Gen-2 RFID system in reference to EPCglobal class1 Gen 2 specifications. We introduce the encoding and backscatter modulation process as well as the BER performance of FM0 Code. In addition, the importance and the role of the frequency and time synchronization, such as the preamble and frame synchronization are explained.

REFERENCES

[1] EPCglobal, EPC radio frequency identity protocols classe-1 generation-2 UHF RFID, protocol for communications at 860 MHz 960 MHz, version 1.0.9, 2004.
[2] F. Hussien et al., "Design considerations and trade-offs for passive RFID tags,", Texas A&M University presentation., May 2005.
[3] D. Dobkin, "The RF in RFID", available online at http://www.enigmatic-consulting.com/Communications_articles/RFID/RF_in_RFID_index.html.
[4] Atmel Corporation, "U2270B, Read/Write Base Station," 2003. Available online at http://media.digikey.com/PDF/Data%20Sheets/Atmel%20PDFs/U2270B.pdf.
[5] SeungHwa Ryu, *RFID in ubiquitous society*. The Electronic Times, March 2005 (Korean).

[6] B. Sklar, *Digital Communications*. 2nd Edition. Prentice Hall PTR, 2002.
[7] A. Goldsmith, *Wireless Communications*. Cambridge University Press, 2005.
[8] M. Simon and D. Divsalar, "Some interesting observations for certain line codes with application to RFID," *IEEE Trans. on Communications*, vol. 54, no. 4, pp. 583-586, April 2006.
[9] Atmel Corporation, "E5551, Standard R/W Identifications IC with Anti-collision", Rev. A3, Oct. 2004. Online at : http //www.rfcard.com.cn/information/jszl/E5551_IN/E5551_IN.pdf, p. 15.